\documentclass[conference]{IEEEtran}
\IEEEoverridecommandlockouts

\usepackage{cite}
\usepackage{mdframed}
\usepackage{hyperref}
\usepackage{multicol}
\usepackage{enumitem}
\usepackage{subfigure}
\usepackage{float}
\usepackage{lipsum}
\usepackage{bbm}
\usepackage{url}
\usepackage{amsmath,amssymb,amsfonts}
\usepackage{algorithmic}
\usepackage{graphicx}
\usepackage{textcomp}
\usepackage{xcolor}
\usepackage{csquotes}
\def\BibTeX{{\rm B\kern-.05em{\sc i\kern-.025em b}\kern-.08em
    T\kern-.1667em\lower.7ex\hbox{E}\kern-.125emX}}
\begin{document}

\title{Learning to Incentivise: Using Reinforcement Learning
for Sustainable Urban Mobility\\
\thanks{This research was partially funded by the Research Council of Finland projects AIforLEssAuto (no. 347200) and ALCOSTO (no. 349327), and the EU-funded research project ACUMEN (no. 101103808).}
}

\author{\IEEEauthorblockN{Germán Pardo-González}
\IEEEauthorblockA{\textit{Dept. of Mathematics}\\ 
\textit{London School of Economics and Political Science}\\
London, UK \\
https://orcid.org/0000-0002-6281-1090}\\ 
\IEEEauthorblockN{Katerina Papadaki}
\IEEEauthorblockA{\textit{Dept. of Mathematics}\\
\textit{London School of Economics and Political Science}\\
London, UK \\
k.p.papadaki@lse.ac.uk}
\and 
\IEEEauthorblockN{Shaghayegh Vosough}
\IEEEauthorblockA{\textit{Dept. of Built Environment}\\
\textit{Aalto University}\\
Espoo, Finland \\
shaghayegh.vosough@aalto.fi}\\
\IEEEauthorblockN{Claudio Roncoli}
\IEEEauthorblockA{\textit{Centre for Industrial Management/Traffic and Infrastructure, KU Leuven}\\ 
Leuven, Belgium\\
\textit{Dept. of Built Environment, Aalto University}\\
Espoo, Finland \\
claudio.roncoli@kuleuven.be}
}

\maketitle

{\footnotesize
\noindent\textbf{Published in:} \emph{2025 9th International Conference on Models and Technologies for Intelligent Transportation Systems (MT-ITS)}, IEEE, 2025. \href{https://doi.org/10.1109/MT-ITS68460.2025.11223596}{DOI: 10.1109/MT-ITS68460.2025.11223596}.\par
\medskip
\noindent\textcopyright{} 2025 IEEE. Personal use of this material is permitted. Permission from IEEE must be obtained for all other uses, in any current or future media, including reprinting/republishing this material for advertising or promotional purposes, creating new collective works, for resale or redistribution to servers or lists, or reuse of any copyrighted component of this work in other works.\par
}
\medskip

\begin{abstract}
Traffic management has traditionally focused on toll-based and road pricing solutions. However, road pricing often raises concerns about accessibility and public dissatisfaction, leading to its prohibition in some places, such as Finland. This study optimises the dynamic allocation of incentives to drivers, encouraging them to reroute onto alternative (potentially longer) paths to achieve greater societal benefit, reduced total travel time (TTT) and total emissions in the transportation network, contributing to sustainable urban mobility. We employ a multi-agent reinforcement learning approach to dynamically assign incentives to drivers to reduce both TTT and emissions, with travel times estimated using traffic simulation software. 
We demonstrate that, with an unlimited budget and an objective of minimising travel time, the incentive scheme reduces TTT by 16\%, compared to the dynamic User-Equilibrium (UE) with a budget equivalent to about 11\% of the UE total time. When the goal is to minimise emissions, a 9\% reduction in CO2 emissions is observed under an unlimited budget. We demonstrate a critical trade-off: minimising TTT leads to an increase in emissions, while prioritising emission reductions raises TTT. However, with the right combination of weights in the multi-objective function, both TTT and total emissions are improved beyond the baseline.
\end{abstract}


\section{Introduction}

Traffic congestion in large cities is a major contributor to air pollution, with road transport accounting for nearly 26\% of the UK's total emissions in 2021 \cite{UK}. Stop-and-go traffic and idling vehicles worsen fuel consumption, increasing pollutants such as nitrogen oxides and particulate matter. Although the Net Zero Emissions scenario projects that 60\% of car sales will be electric vehicles by 2030, four out of five cars on the road will still rely on internal combustion engines \cite{IEA}. To tackle these environmental challenges, urban areas have implemented various traffic management strategies, including congestion pricing, as seen in London \cite{TfL}. While such schemes help reduce vehicle numbers in specific zones, they often face opposition due to fairness concerns and regulatory barriers. As an alternative, incentive-based approaches have gained attention for their ability to promote voluntary changes in travel behaviour \cite{trb, luan2023non}.

Incentive schemes have been proposed with the aim of reducing emissions and congestion by influencing travellers' mode, route, or departure time choices. In the case of route decisions, for instance, if one route is faster but heavily congested, while an alternative is longer but produces fewer emissions, travellers may opt for the latter if compensated appropriately. However, financial constraints limit their widespread implementation, making efficient allocation critical. While an unlimited budget could theoretically optimise traffic flow and emissions reduction, real-world applications require strategic distribution of incentives to achieve the greatest environmental benefits while maintaining smooth traffic movement.

To minimise emissions while achieving SO traffic flow, some drivers must take routes that are (slightly) longer than their shortest paths \cite{van2019c}. To facilitate this shift, incentive schemes can be implemented to encourage drivers to opt for routes that may be less desirable in terms of personal travel time \cite{Vosough2024TP,klein2018emergence,djavadian2014empirical}, but contribute to higher overall social benefits. While numerous studies have analysed the effects of incentive schemes on total travel time and emissions under static traffic conditions \cite{hEART,luan2023non}, the dynamic nature of traffic must be considered to effectively reduce emissions. This highlights the necessity of designing incentive mechanisms within a dynamic framework \cite{hEART2025}.

While no studies have directly examined incentive schemes for emissions reduction in urban transportation networks, some research has employed optimisation-based approaches to regulate tolls for traffic management with the goal of reducing emissions. For example, \cite{toll1} developed a bi-level optimisation model where the lower level determines an equilibrium traffic assignment, while the upper level applies different tolling strategies specifically designed to minimise CO2 emissions. However, their toll strategies are categorical and may yield suboptimal solutions. On the other hand, \cite{RL3} formulated a route-based traffic assignment approach incorporating tolls determined via Multi-agent Reinforcement Learning (MARL). Their tolling mechanism utilises a reward function that balances route travel time and imposed tolls, allowing users to prioritise these objectives.

The objective of this study is motivated by the incentive schemes designed in~\cite{hEART, hEART2025}, where the authors investigated the effectiveness of static link and path incentive schemes on the total travel time (TTT) under a budget limit. Our work extends their approach to a dynamic setting using simulation and employing a multi-objective approach, introducing the estimation of emissions and insights shown in a Pareto front for different budgets.

Building on top of \cite{hEART2025}, this study introduces a multi-objective reward function that simultaneously minimises congestion and emissions. The latter are estimated using the data given by the microscopic traffic simulator SUMO, namely the acceleration and speed of the vehicle, and we examine the trade-off between travel efficiency and environmental impact. By generating a Pareto front under different budgets, we provide insights into the balance between incentive allocation and system-wide improvements. The proposed approach is evaluated against a baseline obtained from SUMO that uses the Gawron's algorithm for dynamic equilibrium \cite{simulationbased}, using the Helsinki city centre (Kamppi area) network as a case study, and we derive the weights (or objective priorities) that yield better performance (in terms of travel time and emissions) than the baseline.

Our main contributions are:

\begin{enumerate}
\item Developing a multi-objective MARL-based incentive scheme to minimise TTT and CO2 emissions simultaneously.
\item Estimating emissions through data from the SUMO simulator and analysing the Pareto-optimal trade-off between congestion and emissions under various incentive budgets.
\item Comparing the proposed scheme to the SUMO baseline, demonstrating its effectiveness in improving network efficiency while reducing environmental impact in a real-world urban setting.
\end{enumerate}

The remainder of this paper presents the methodology employed to benchmark and investigate the incentives scheme in Section~\ref{Methodology}; the case study, outputs of the algorithms and the results’ interpretations in Section~\ref{Results}; and finally, conclusions are drawn in Section~\ref{Conclusions}.

\section{Methodology} \label{Methodology}

In this section, we formulate a Markov Decision Process (MDP) using the universal modelling framework from \cite{powell1}. This approach extends our work~\cite{hEART2025}, incorporating emissions in a multi-objective reward function. Based on MARL and microscopic simulation techniques, our model aims to determine the optimal incentive for each driver under a limited budget. We solve this MDP employing the Independent Q-learning algorithm \cite{MARL}, an algorithm adapted for multi-agent environments where each traveller, defined by their origin and destination, acts as an agent with a corresponding Q-function. Additionally, we detail our methodology for estimating CO2 emissions and integrating these estimates into the multi-objective reward function.

\subsection{Markov Decision Process Formulation}
\textit{State Variable.} 
The state represents all the information needed to decide what happens after an action is taken. The goal is to find an optimal policy, i.e., a set of rules that tells the agent the best action to take. In this case, the \enquote{state} of an agent is simple: their origin and destination. Once an agent selects a route (an action), it reaches its destination without perturbations, and thus, it is not necessary to store the state. This is called a one-step MDP because the agent's decision is made at the start, and there are no further transitions or decisions along the way.
\\

\textit{Decision Variable.} 
The (reinforcement learning) agent should decide whether a path should be incentivised or not, with the restriction that at most one can be incentivised. Note that the agent could have the decision that no paths are incentivised. Let $\mathcal{W}$ be the set of agents and $\mathcal{P}_w$ the set of available paths for agent $w\in\mathcal{W}$. Also, for a given agent $w\in\mathcal{W}$, $X_{w}$ is a vector in $\mathbb{R}^{|\mathcal{P}_w|}$ with binary entries that represent the taken path $p\in\mathcal{P}_w$. Moreover, the set that encompasses the paths taken by all agents is $X=\{X_{w}\}_{w\in \mathcal{W}}$.

Then, our decision variable is a vector $Y_{w}\in \mathbb{R}^{|\mathcal{P}_w|}$ with binary entries where 1 means that the path will be incentivised and 0 otherwise. Consider the action space, which represents the set of all possible decisions for agent $w\in\mathcal{W}$:
$$
\mathcal{Y}_{w} \in \left\{ y_{w}\in\{0,1\}^{|\mathcal{P}_w|}  : y_w\mathbbm{1} \leq 1 \right\}, \quad \forall w \in \mathcal{W}
$$
for which $Y_w$ takes values in $\mathcal{Y}_w$. 

Note the connection between $Y$ and $X$, as the latter gives the paths used by all agents. Even though the decision variable is $Y_w$, we still use $X_w$ and $X$ to calculate the reward. In case one path is incentivised, it will be used, and thus $X_w=Y_w$. However, when $Y_w=0$ (no path is incentivised), the path taken corresponds to the shortest.

Let the vector of path travel times of all paths of agent $w\in\mathcal{W}$ be $\tau_w\in \mathbb{R}^{|\mathcal{P}_w|}$. Let the scalar $\delta$ be the amount we reduce the time of the incentivised path. Thus, the travel times are adjusted as follows.
$$
\tau_w \leftarrow \tau_w- Y_w  \delta. 
$$
The above update only modifies a single entry of $\tau_w$, the one that corresponds to the incentivised path. Since we want the incentivised path to have the lowest travel time, we set the time reduction $\delta$ to be the difference between the incentivised path's travel time and the minimum path's travel time, plus an extra scalar term $\phi>0$:

$$
\delta = \tau_w^T Y_w - \min\{\tau_w\} + \phi.
$$
This will make the incentivised path's travel time to be $\phi$ units below the minimum travel time.

Finally, the path with minimum cost is selected, which corresponds to the incentivised one, namely, $\arg\min_{p\in\mathcal{P}_w}\{\tau_w\}$. If there are no incentivised paths, all entries of $Y_w$ will be 0, and no modification will be made.
\\

\textit{Budget Limitation.} Let $B$ be the total budget available for incentivising drivers and $b$ the budget used so far. For convenience, the budget is measured in time units. The algorithm checks whether there is enough budget to complete the action, i.e., $\delta + b \leq B$. If there is an insufficient budget, it does not modify the travel times and selects the original shortest path. The budget is continuously tracked throughout the process, representing the state of the central authority responsible for assigning incentives.
\\

\textit{Reward Function.} 
The reward function consists of two main terms, $R^\textrm{T}_{w}(X)$ and $R^\textrm{E}_{w}(X)$, associated with time and emissions, respectively. We first show the reward function for travel time:
$$
    R^\textrm{T}_{w}(X) =  \omega_1 C_w(X) + \omega_2 \text{TTT}(X), \quad \forall w\in\mathcal{W}
$$
where $\omega_1$ and $\omega_2$ are chosen to ensure that the two quantities are of the same order of magnitude. Each vehicle is treated as an autonomous agent, with specific characteristics, such as velocity $v$ and acceleration $a$.
These rewards correspond to a weighted combination of the agent's travel time and the TTT, which are known only after all agents have made their routing decision and the behaviour of each individual driver on the network has been implemented. We define $C_w(X)$ to be the travel time experienced by agent $w\in\mathcal{W}$, given the choice of the other agents stored in $X$. Also, TTT$(X)$ is the estimated total travel time of the network that also depends on the actions of every agent.

The reward function for emissions is defined as:
$$
    R^\textrm{E}_{w}(X) =  \lambda_1 Em_w(X) + \lambda_2 \text{TE}(X), \quad \forall w\in\mathcal{W}
$$
where $\lambda_1$ and $\lambda_2$ are chosen to ensure that the two quantities are of the same order of magnitude, as above. Similarly, these rewards correspond to a weighted combination of the agent's individual emissions and the total emissions. We define $Em_w(X)$ to be the generated emissions by agent $w\in\mathcal{W}$, given the choice of the other agents stored in $X$. Also, TE$(X)$ are the estimated total emissions of the network that also depend on the actions of every agent. The emissions are calculated based on the dynamic model in \cite{emissions_calc}, with the necessary parameters, namely, the acceleration and velocity of each vehicle, obtained from SUMO.

We finally show the complete reward function $R_w(X)$ as follows:
\begin{equation}\label{eqReward1}
    R_{w}(X) =  R^\textrm{T}_{w}(X) + R^\textrm{E}_{w}(X), \quad \forall w\in\mathcal{W}
\end{equation}
Recall that to compute the reward, we need to know $X$, as it gives the paths used by all agents. In fact, even though the decision variable is $Y_w$, $X_w$ and $X$ are used to calculate the reward: in case one path is incentivised, it will be used, and thus $X_w=Y_w$; however, when $Y_w=0$ (no path is incentivised), then $X_w$ will have entry $1$ for the path given by $\arg\min_{p\in\mathcal{P}_w}\{\tau_w\}$, which is the shortest path, i.e., the used one. 
\\

\textit{Emissions Estimation.}
We build on top of the  Newton-based greenhouse gas model dynamic model~\cite{emissions_calc} to estimate emissions generated by each agent $w\in\mathcal{W}$ based on their acceleration, $a_w$, and velocity, $v_w$. Table ~\ref{params} depicts the definitions and values of all the relevant parameters used in this model. 
The emissions are calculated as
\begin{equation}\label{emissions_eq}
Em_{w} =\begin{cases}
    r & \textrm{if } \gamma_w \leq 0
    \\
    \gamma_w (v_w + 0.5 a_w) / \eta&\text{otherwise}
\end{cases}
\end{equation}
where
\begin{align}
    \gamma_w &= \frac{T_\text{idle}}{E_\text{gas}} \zeta_w \\
    \zeta_w &= M a_w v_w + M g c_{rr} v_w * 0.5 c_d A \rho v_w^3.
\end{align}

Finally, the total emissions are calculated as $\text{TE}=\sum_{w\in\mathcal{W}}Em_{w}$, i.e., the sum of the emissions generated by all the vehicles.

\begin{table}[tb]
    \centering
    \caption{Newton-based greenhouse gas model parameters.}
    \begin{tabular}[t]{ccc}
    \hline
    Parameter & Definition & Value\\
    \hline
     {$T_\text{idle}$} & CO2 emissions from gasoline & 8887 gCO2/gal\\
     {$M$} & Vehicle mass & 1334 kg\\
     {$a_w$} & Vehicle acceleration & From SUMO $[m/s^2]$\\
     {$v_w$} & Vehicle speed & From SUMO $[m/s]$\\
     $g$ & Gravitational acceleration & 9.81 $[m/s^2]$\\
     $c_{rr}$&Rolling resistance & 0.015 \\
     $c_d$ & Aerodynamic drag coefficient & 0.3\\
    $\eta$ & Fuel efficiency & 0.7\\
     $A$ & Frontal area & 2.5 $[m^2]$\\
     $\rho$ & Air density &  1.225 $[kg/m^3]$\\
     $E_\text{gas}$ & Energy in gas & $31.6\times10^6$ [J/L] \\
     $r$ & Regeneration efficiency ratio & 0\\
    \hline
    \end{tabular}
    \label{params}
\end{table}

\textit{Objective Function.}
We consider the classical action-value function (Q-function) which corresponds to the value of taking a given action while being in a given state. But, as we do not store the state (as the state remains unchanged), this reduces to an action function that represents the value of taking an action regardless of the state. We use a different Q-function for each agent:
\begin{equation}
    Q_w^*(Y_w) = \min_{X'\in\mathcal{X}} R_{w}(X').
\end{equation}

The above equation is solved iteratively for every episode $n$ using the following update rule.
\begin{equation}\label{eqUpdates2}
Q_w(Y_w) \leftarrow (1-\alpha_n)Q_w(Y_w) + \alpha_n  R_{w}(X), \qquad \forall w\in\mathcal{W}
\end{equation}
where $\alpha_n$ is the learning rate, that influences the learning performance of the algorithm. One typically starts with a big $\alpha_n$ so that the algorithm gives more importance to new information. At the end, we want a small $\alpha_n$ so that it converges. 

\subsection{Solution Algorithm.}
We employ the following algorithm to solve our problem.

{\setlength{\parindent}{15pt}
\begin{mdframed}
    \textbf{Initialise}
    \begin{enumerate}
        \item Compute $k$ shortest paths for every agent $w \in \mathcal{W}$ and store them in $\mathcal{P}_w$.
        \item For each $w \in \mathcal{W}$, set the entries of $\tau_w$ to be the free-flow travel times of $w$'s chosen paths from SUMO. 
        \item Set $Q_w(Y_w) \leftarrow 0$ for all agents $w \in \mathcal{W}$ and set $\phi$ as a small number.
    \end{enumerate}
    \textbf{For each episode $n = 1, ..., N$:}
    \begin{enumerate}[resume]
        \item Set $\epsilon_n = 1-\frac{n}{N}$, $\alpha_n = \frac{a}{(b + n)}$ and $b=0$.
    \end{enumerate}
    \hspace{13pt}\textbf{For each agent $w\in\mathcal{W}$:} 
    \begin{enumerate}[resume, leftmargin=3\parindent]
        \item Sample random number $p\sim U(0,1)$.
        \item Take action $Y_w$ as follows:
        
        $Y_w =\begin{cases}
            \text{Random action } Y_w \in \mathcal{Y}_w, &
            \textrm{if } p \leq \epsilon_n
            \\
            \text{arg}\min_{Y_w \in \mathcal{Y}_w} Q_w(Y_w), &\text{otherwise}
        \end{cases}$
        \item Calculate the amount of the incentive $\delta$: 
            $$\delta = \tau_w^T Y_w - \min\{\tau_w\} + \phi$$
            
        \item If there is enough budget left, i.e., $\delta + b \leq B$:
            \begin{itemize}
                \item[] Modify travel times $\tau_w$ and update budget $b$:
                $$\tau_w \leftarrow \tau_w - Y_w \delta$$
                $$b \leftarrow b + \delta$$
            \end{itemize}
            Else:
            \begin{itemize}
                \item[] Do not apply incentives.
            \end{itemize}
            
            \item Select the path to be taken: $p =\arg\min_{p \in \mathcal{P}_w}\{\tau_w\}$. Let $X_w$ be the decision vector of choosing path $p$.
            
            \item Add $X_w$ to $X$, the set of all paths taken by agents.
    \end{enumerate}
    \begin{enumerate}[resume]
        \item After all agents have completed their trips, retrieve $\text{TTT}(X)$, travel time $C_p(X)$ from SUMO, and update the travel times $\tau_w$ for each agent.
    \end{enumerate}
    \hspace{13pt}\textbf{For each agent $w\in\mathcal{W}$:} 
    \begin{enumerate}[resume, leftmargin=3\parindent]
        \item Compute the reward using Eq. (\ref{eqReward1}), and update the Q-function with Eq.(\ref{eqUpdates2}).
    \end{enumerate}
\end{mdframed}
}

\section{Numerical experiments} \label{Results}

\subsection{Case study}

\begin{figure}[tb]
    \centering
    \includegraphics[width = 2in]{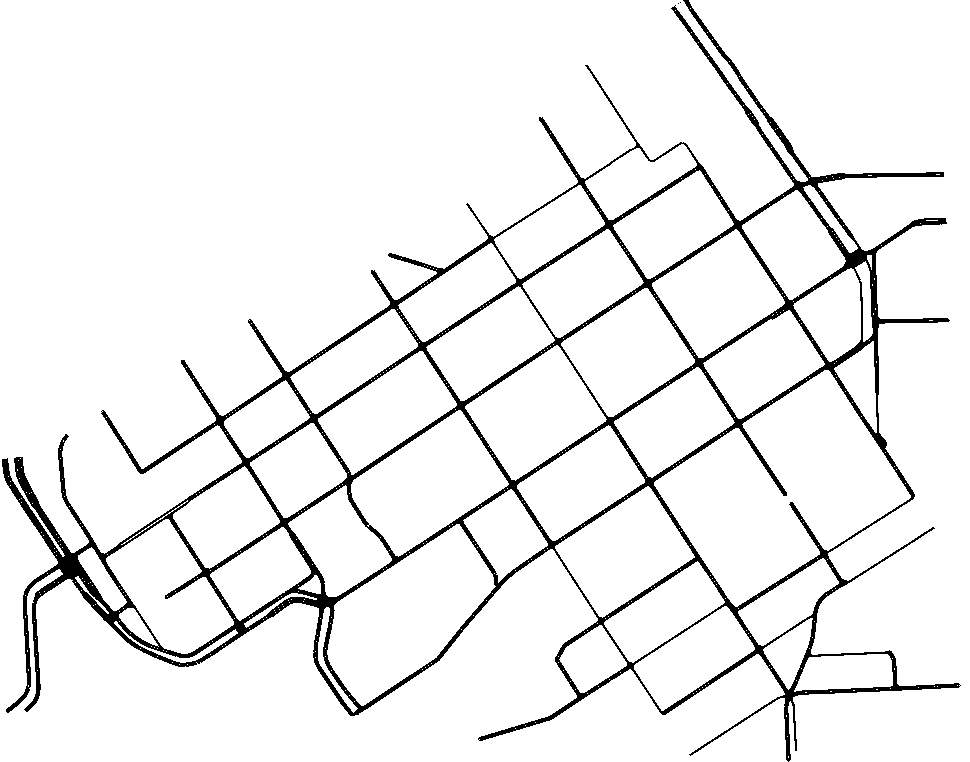}
    \caption{Kamppi, Helsinki Area network.}
\label{kamppi}
\end{figure}

We apply our proposed methodology to the Kamppi area, Helsinki, Finland, on the network depicted in Fig. \ref{kamppi}. This area is known for its congestion during peak hours and its central location.

The data for simulation in SUMO, including a set of trips with departure times for each traveller and the network definition, is obtained from \cite{ODpairs}. The network contains 235 edges and 152 nodes, and there are 1100 O-D pairs.

\subsection{Numerical results}

\begin{figure}[tb]
    \centering
    \subfigure[]{
    \includegraphics[scale=0.45]{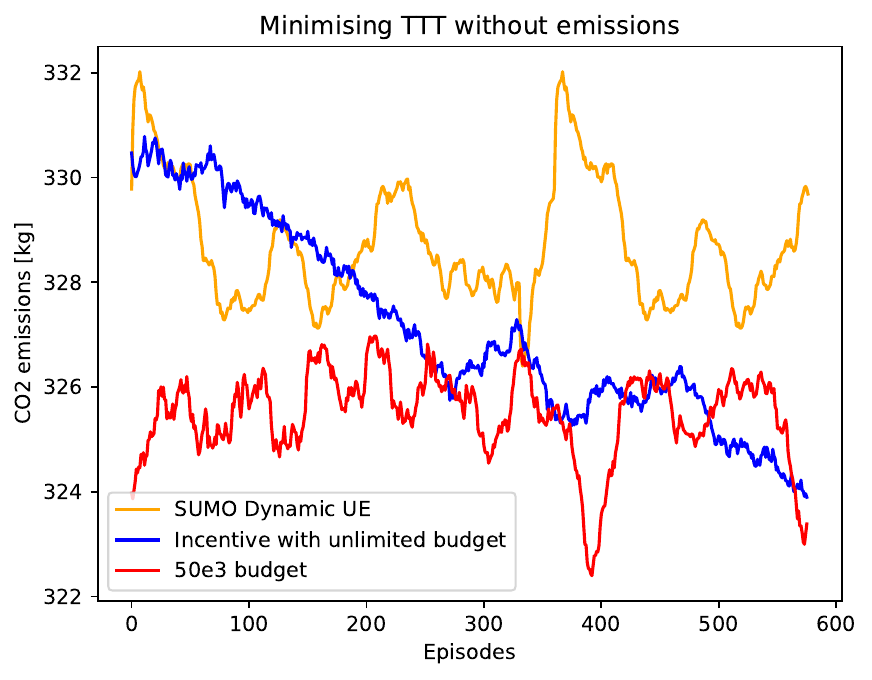}
    }
    \subfigure[]{
    \includegraphics[scale=0.45]{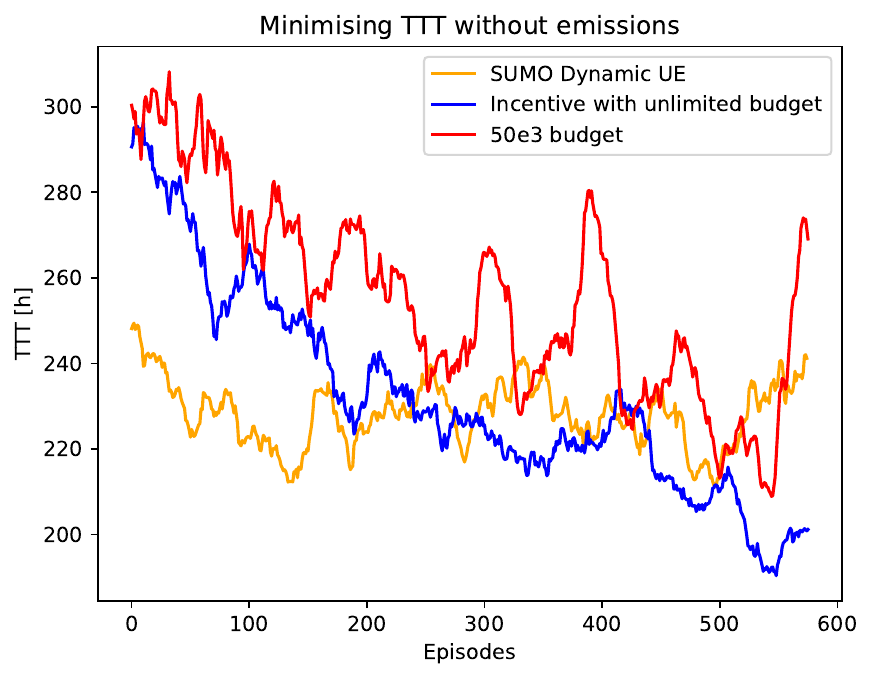}
    }
    \caption{CO2 and TTT when minimising TTT under different incentive budgets.}
\label{TTT_fig}
\end{figure}

\begin{figure}[tb]
    \centering
    \subfigure[]{
    \includegraphics[scale=0.45]{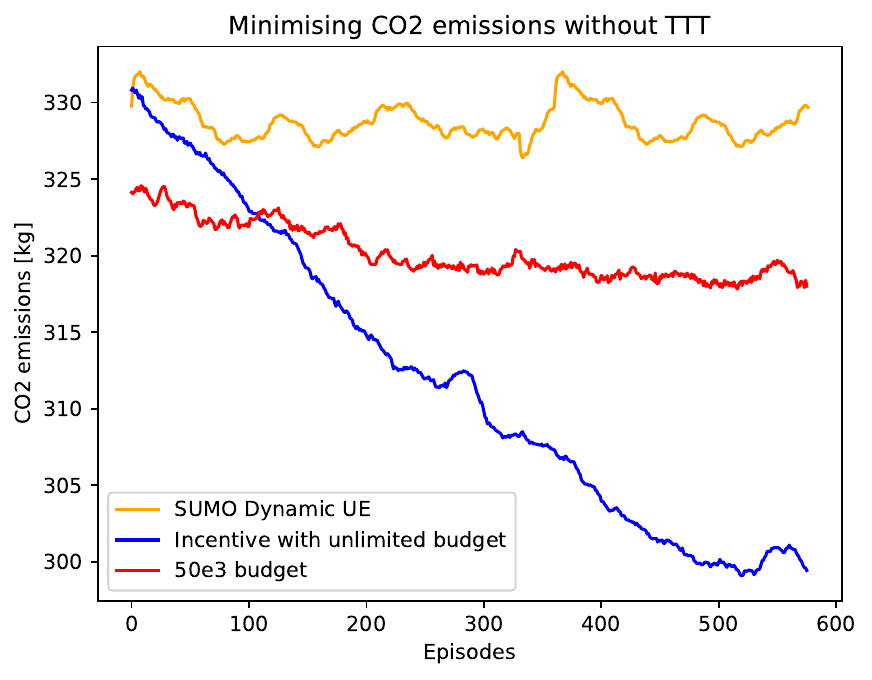}
    }
    \subfigure[]{
    \includegraphics[scale=0.45]{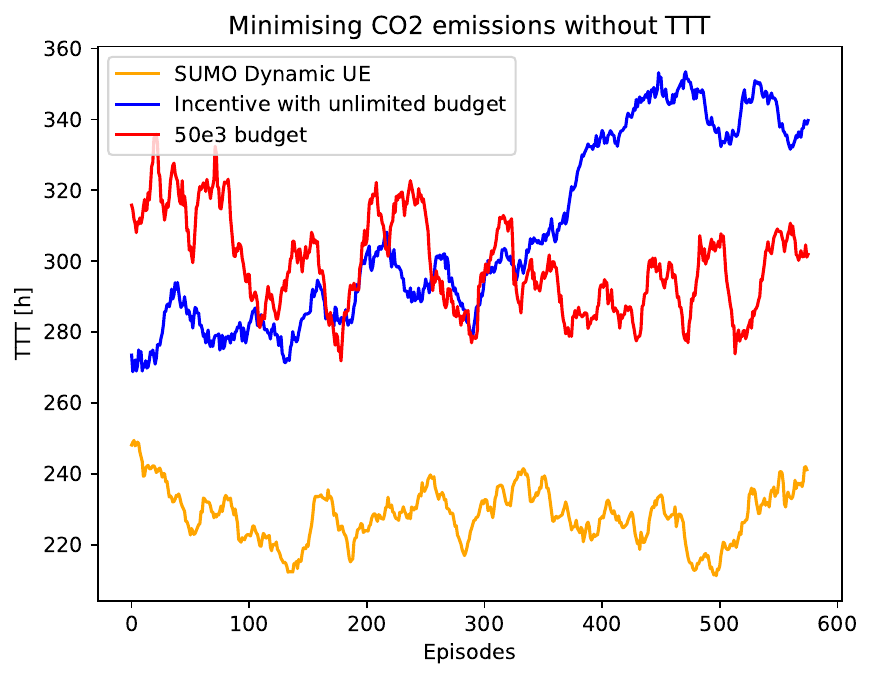}
    }
    \caption{CO2 and TTT when minimising emissions under different incentive budgets.}
\label{emissions_fig}
\end{figure}
\begin{figure}[tb]
    \centering
    \subfigure[]{
    \includegraphics[scale=0.45]{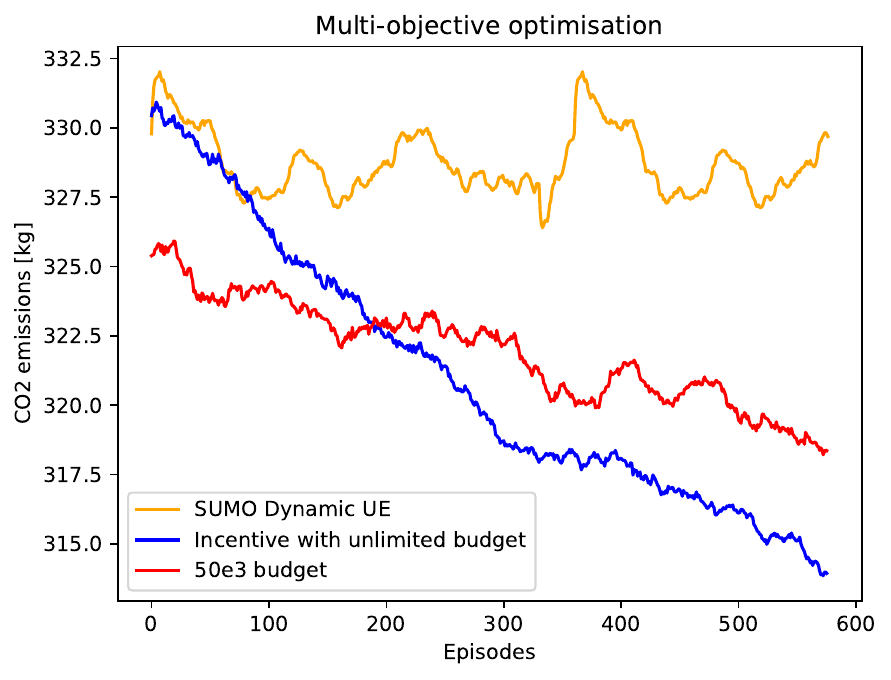}
    }
    \subfigure[]{
    \includegraphics[scale=0.45]{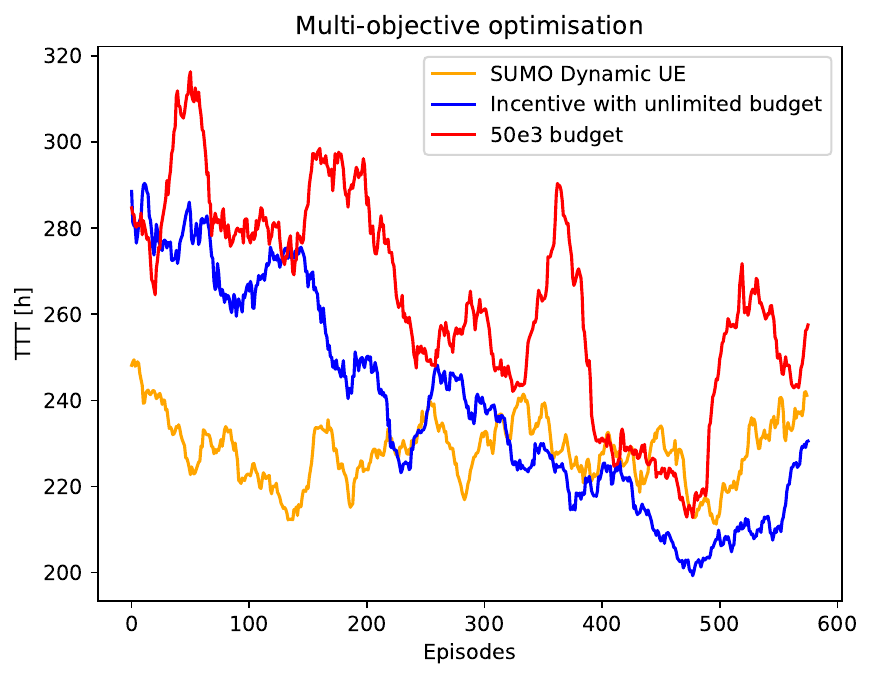}
    }
    \caption{CO2 and TTT when mixing travel time and emissions in the objective under different incentive budgets.}
\label{multi_obj}
\end{figure}
\begin{figure}[tb]
    \centering
    \includegraphics[scale=0.3]{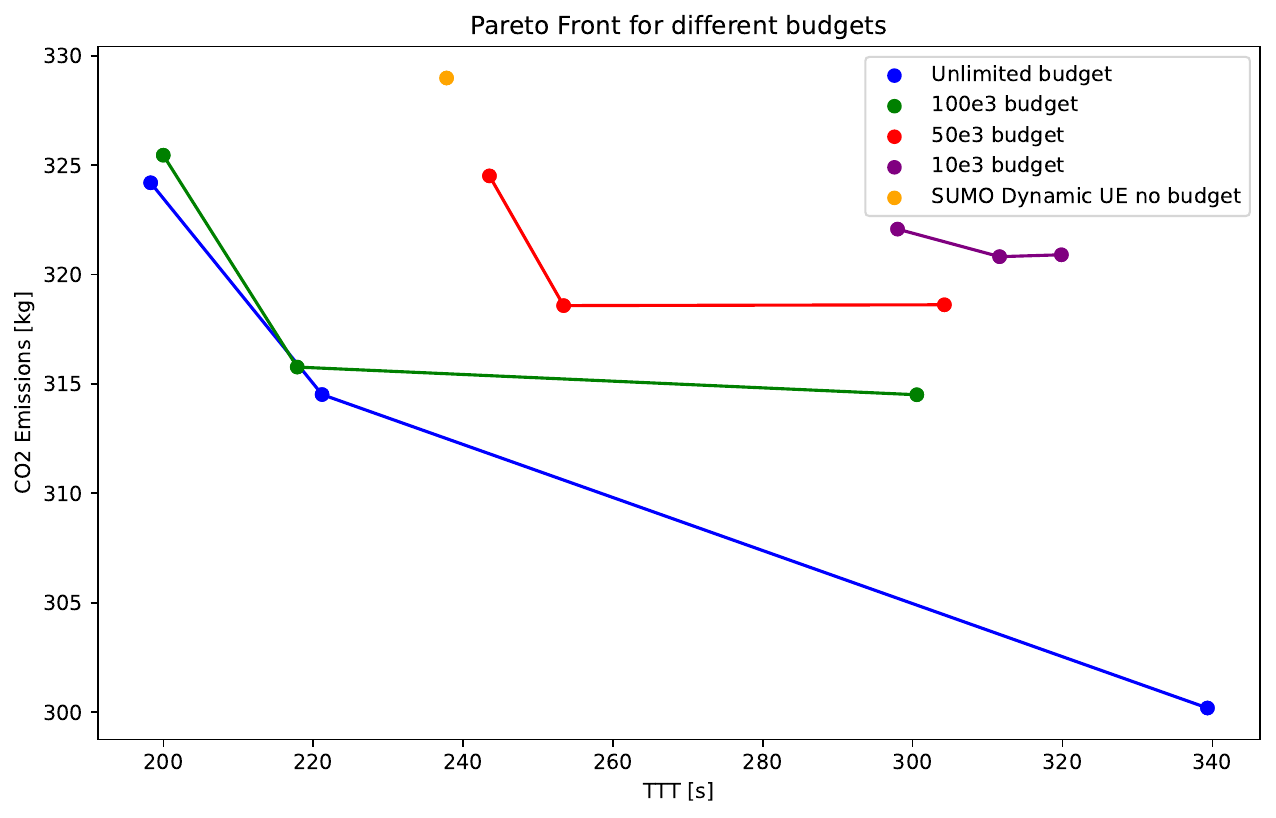}
    \caption{Pareto fronts for different incentive budgets.}
\label{pareto}
\end{figure}

In this study, we demonstrate the trade-off between optimising for travel time or optimising for emissions generated over the network. As a first example, Fig.~\ref{TTT_fig} shows the behaviour of the CO2 emissions and the TTT when the objective is to minimise travel time, i.e., $\omega_1, \omega_2\geq0$. The TTT outperforms the baseline by SUMO, and as the budget decreases, the solution degrades. Regarding the emissions, these are reduced for an unlimited budget, and our methodology outperforms the baseline for any budget.

On the other hand, Fig.~\ref{emissions_fig} shows the results when the objective is set to minimise emissions only, i.e.,~$\lambda_1,\lambda_2\geq0$. From the figures that are minimising emissions, the reader may appreciate the clear difference between the budget allowance and the dynamic UE baseline for emissions. It is clear how the emissions generated are lowest with an unlimited budget, and the solution degrades as the budget allowance decreases, while always outperforming the dynamic UE baseline. Regarding the TTT, there is a negative trend as it seems to increase (even greater than the baseline) when the emissions are minimised. This suggests that in order to minimise emissions and maintain a reasonable TTT, both objectives should be taken into account in the reward function, giving place to a multi-objective optimisation problem.

Regarding the multi-objective reward function, we analyse the results giving the same weight to TTT and emissions. When using this combination, the results in Fig.~\ref{multi_obj} were obtained, where we can notice how the emissions are significantly lower than the dynamic UE baseline, but not as low as in Fig.~\ref{TTT_fig}, which intuitively reasons as the objective now considers both travel time and emissions and these are conflictive. Conversely, it is more difficult to appreciate the time benefit, as the incorporation of emissions seems to degrade the TTT. With an unlimited budget, the algorithm ends up converging with a lower TTT, but for a budget of 50e3, it appears slightly worse at the end. This suggests that, for a small budget, the weights corresponding to time should be higher than the ones corresponding to emissions, so that the algorithm prioritises time, resulting in a scenario where both emissions and TTT are better than the baseline. 

Finally, and most importantly, in relation to the multi-objective reward function, we have identified the conflicting nature of emissions and travel time. The Pareto front of non-dominated solutions is shown in Fig.~\ref{pareto} for different budgets and the dynamic UE baseline. The curves illustrate how the solution deteriorates as the budget decreases. They also explain why the TTT is not as low when the budget is set at 50e3 and provide insights into how the weights of the objective function can be adjusted to outperform the baseline in both emissions and TTT. Notably, our approach consistently outperforms the baseline in terms of emissions. However, when considering TTT, a certain budget allowance is required to achieve superior performance. The transition to net-zero carbon emissions is more readily attainable, as emissions are minimised most effectively. However, we aim to avoid a significant increase in TTT. Therefore, the weights should be selected in a way that heavily minimises emissions while maintaining an acceptable TTT.

\section{Conclusions} \label{Conclusions}
We implement a dynamic incentives scheme aimed at reducing both emissions and overall congestion in the network. With this, we demonstrate how encouragement can lead to social benefits by minimising emissions, TTT and a mix of both while offering a more publicly acceptable alternative to road pricing, which often causes public dissatisfaction and uneven welfare distribution.

To this end, we extend the work in \cite{hEART2025} to appropriately calculate and incorporate emissions into the algorithm and consider a multi-objective reward function. The solution algorithm is based on MARL, where SUMO is the environment to estimate travel times dynamically for more realistic results, and we use simulated data from the Kamppi area in Helsinki, Finland. The results indicate that our algorithms are reliable for an unlimited budget and show what objective should be prioritised to outperform the dynamic UE baseline. 

Some of the limitations of this work include not considering the behaviour of the drivers, as in realistic situations, a driver may follow or not the incentivised route. This can be taken into account by assuming a compliance rate. Furthermore, as shown in \cite{hEART}, incentives on paths outperform incentives on links. None of them were tried in this study, as the incentives are along trips; however, it would be insightful to compare results with path incentives.

\section*{Acknowledgment}
This work was partially done during the MSc of Germán Pardo-González at the London School of Economics and Political Science. The first author extends gratitude to Aalto University for its hospitality and for providing the opportunity to work on this study through the Aalto Science Institute (AScI) internship program. The work of Shaghayegh Vosough and Claudio Roncoli was partly funded by the EU-funded research project ACUMEN (no. 101103808).

\bibliographystyle{IEEEtran}
\bibliography{ref}
\end{document}